\documentclass[11pt,a4paper]{article}
\usepackage[utf8]{inputenc}
\usepackage{times}
\usepackage[T1]{fontenc}
\usepackage[english]{babel}
\usepackage[pdftex]{graphicx}
\usepackage{xcolor}
\usepackage{a4}
\usepackage{amsfonts,amsmath,amssymb,bbm,slashed}
\usepackage[amssymb,Gray]{SIunits}
\usepackage{url}
\usepackage[numbers,sort&compress]{natbib}
\usepackage[normalem]{ulem}
\usepackage[a4paper]{geometry}
\usepackage{graphicx}
\usepackage[margin=\parindent,font=small,labelfont=bf]{caption}
\usepackage[caption=false]{subfig}
\renewcommand\appendix{\par 
  \setcounter{section}{0}%
  \setcounter{subsection}{0}%
  \setcounter{figure}{0}%
  \renewcommand\thesubsection{\Alph{subsection}}%
  \renewcommand\thefigure{\Alph{subsection}.\arabic{figure}}
  \numberwithin{equation}{subsection}}
\newcommand{\abs}[1]{\ensuremath{\left\vert #1 \right\vert}}

\newcommand{\bleq}{\ensuremath{\mathrel{\phantom{=}}}}

\newcommand{\bra}[1]{\langle #1 \hspace{-2pt} \mid}
\newcommand{\ket}[1]{\mid \hspace{-1pt} #1 \rangle}
\newcommand{\D}{\mathrm{d}}

\newcommand{\sne}{Schr{\"o}\-din\-ger--New\-ton equation}
\newcommand{\schr}{Schr{\"o}\-din\-ger}
\newcommand{\abbrv}[1]{\begin{small}#1\end{small}}

\renewcommand{\vec}[1]{\mathrm{\mathbf{#1}}}

\title{The Schr\"odinger--Newton equation and its foundations}
\author{Mohammad Bahrami, Andr\'e Gro{\ss}ardt, Sandro Donadi, and Angelo Bassi \\
\footnotesize\textit{Department of Physics, University of Trieste, 34151 Miramare-Trieste, Italy}\\
\footnotesize\textit{Istituto Nazionale di Fisica Nucleare, Sezione di Trieste, Via Valerio 2, 34127 Trieste, Italy}\\
\footnotesize\textit{Electronic mail addresses: first name.last name@ts.infn.it}}

\begin{document}

\maketitle

\begin{abstract}\noindent
The necessity of quantising the gravitational field is still subject to an open debate.
In this paper we compare the approach of quantum gravity, with that of a fundamentally semi-classical theory
of gravity, in the weak-field non-relativistic limit. We show that, while in the former case the \schr\
equation stays linear, in the latter case
one ends up with the so-called \sne, which involves a nonlinear, non-local gravitational contribution.
We further discuss that the \sne\ does not describe the collapse of the wave-function, although it was initially
proposed for exactly this purpose.
Together with the standard
collapse postulate, fundamentally semi-classical gravity gives rise to superluminal signalling.
A consistent fundamentally semi-classical theory
of gravity can therefore only be achieved together with a suitable prescription of the wave-function collapse.
We further discuss, how collapse models avoid such superluminal signalling
and compare the nonlinearities
appearing in these models with those in the \sne.
\end{abstract}

\section{Introduction}
The \sne,
\begin{equation}
\label{eqn:sn}
\mathrm{i} \hbar \partial_t \psi(t,\vec r) = \left( -\frac{\hbar^2}{2 m} \nabla^2
- G m^2 \int \D^3 \vec r' \, \frac{\abs{\psi(t,\vec r')}^2}{\abs{\vec r - \vec r'}}
\right) \psi(t,\vec r) \,,
\end{equation}
has been brought into play by Di{\'o}si~\cite{Diosi:1984} and Penrose~\cite{Penrose:1996,Penrose:1998,Penrose:2014}
to provide a dynamical description of the collapse of the quantum wave-function.
It has regained attention in recent times, mainly due to its connection with the question whether gravity
should really be quantised~\cite{Carlip:2008}, and due to its falsifiability in envisaged
experiments~\cite{Giulini:2011,Giulini:2013,Yang:2013}.

There has been quite a debate about how the \sne\ relates to established fundamental
principles of physics \cite{Anastopoulos:2014,Anastopoulos:2014a,Adler:2007,Christian:1997}.
Here we wish to clarify the motivations underlying the \sne, in particular, in which sense it
follows from semi-classical gravity. We will also discuss its present
limitations as a fundamental description of physical phenomena.

First of all one should avoid confusion about what is actually meant by the
term \textit{semi-classical gravity}. Most physicists assume that the gravitational field must be
quantised in some way or another, and that semi-classical gravity is only an \emph{effective}
theory, which holds in situations where matter must be treated quantum mechanically, but gravity
can be treated classically (although it is fundamentally quantum). In this case, of course, we have a
pure quantum theory where everything remains linear.
From such a theory we would never expect any nonlinear interaction terms in the
\schr\ equation\footnote{The self-interactions present in the classical theory would then be treated in the same
way as in Quantum Electrodynamics, namely, through the normal ordering and renormalisation prescription.
They would lead to a mass-renormalisation of the theory rather than a potential term in the \schr\ 
equation~\cite{Anastopoulos:2014,Anastopoulos:2014a}.}.
Under these assumptions, the \sne\ can be derived
within quantum gravity\footnote{The term \emph{quantum gravity} here is restricted to any such theory which,
at least in its low energy limit, treats the gravitational field in the linearised Einstein equations
as a linear quantum operator.} as a mean-field limit---but only as such---whose validity is restricted
to the case of large numbers of particles; we will review this in section \ref{sec:hartree}.\footnote{%
Apart from this, the \sne\ also follows uncontroversially from a gravitating \emph{classical} matter
field~\cite{Giulini:2012}.}

The point taken by the proponents of equation \eqref{eqn:sn}, however, is different.
The \sne\ does follow from a theory in which only
matter fields are quantised, while the gravitational field remains classical {\it even at the fundamental level}.
This is what we will refer to as \emph{semi-classical gravity} in the following, thereby adopting the
notation of \cite{Kibble:1981,Mattingly:2005,Kiefer:2007}.
One possible candidate for such a theory is a coupling of gravity to matter by means of the
semi-classical Einstein equations
\begin{equation}
\label{eqn:sce}
R_{\mu \nu} + \frac{1}{2} g_{\mu \nu} R = \frac{8 \pi G}{c^4} \,
\bra{\Psi} \hat{T}_{\mu \nu} \ket{\Psi} \,,
\end{equation}
that is by replacing the classical energy-momentum tensor in Einstein's equations by the expectation value
of the corresponding quantum operator in a given quantum state $\Psi$. This idea has a long history, dating
back to the works of M\o{}ller~\cite{Moller:1962} and Rosenfeld~\cite{Rosenfeld:1963}.
It has been commented repeatedly that such a theory would be incompatible with established principles of
physics~\cite{Eppley:1977,Page:1981} but these arguments turn out to be
inconclusive~\cite{Mattingly:2005,Mattingly:2006,Kiefer:2007,Albers:2008}. We review them in appendix~\ref{app:semiclass-arguments}.

Note that the validity of \eqref{eqn:sce} at the
fundamental level requires that the collapse does not violate local energy-momentum conservation,
$\partial^\mu \bra{\Psi} \hat{T}_{\mu \nu} \ket{\Psi} = 0$. 
This is certainly not the case for the standard instantaneous collapse in quantum mechanics.
Also collapse models~\cite{Ghirardi:1986,Ghirardi:1990a,Adler:2007b,Adler:2009,Bassi:2003,Bassi:2013}, that
have been constructed to date, violate this condition. There is, however, no obvious reason why local
energy-momentum conservation must be violated by any measurement prescription. Indeed Wald~\cite{Wald:1994}
shows that such a measurement prescription which is consistent with the semi-classical Einstein equations is possible.

At the current state of physics, the honest answer to the question
if the gravitational field must be quantised is therefore that we do not know. The final answer can only be
given by experiment.
In this regard, it is worthwhile noting that the collapse of the wave function---if it is a real
phenomenon---can only be explained by a nonlinear, i.\,e. non-quantum, interaction.
Therefore, if gravity is responsible for the collapse, as often suggested in the
literature~\cite{Diosi:1987,Diosi:1989,Diosi:2007,Adler:2014}, it must remain fundamentally classical
(or, in any case, non-quantum), but the form of coupling to quantum matter is of course open to debate.

Given that gravity is fundamentally classical, and that equation \eqref{eqn:sce} is a fundamental equation of
nature describing gravity's coupling to matter, and not an effective equation,
the \sne\ follows naturally. We show this in section~\ref{sec:sne_from_sce}. However, it is wrong to jump
to the conclusion that the \sne\ alone represents a coherent description of physical phenomena at the
non-relativistic level. As we explain in section~\ref{sec:collapse}, some collapse rule, or collapse dynamics,
must be added in order to account for the stochastic outcomes for measurements of superposition states.
But even by adding the standard collapse postulate to the \sne\ there are difficulties.
Quantum non-locality (which is implicit in the collapse postulate  together with the \schr--Newton dynamics)
leads to the possibility of faster-than-light signalling, as we will show in section~\ref{sec:thought-experiment}.
In section~\ref{sec:comparison} we compare the \sne\ with the typical equations used in collapse models,
concluding that both types of equations are of a very different structure and cannot easily be combined.

\section{The \sne\ from semi-classical gravity} \label{sec:sne_from_sce}

As anticipated, here we take the point of view that at the fundamental level quantum theory is coupled to
classical General Relativity via the semi-classical Einstein equations~\eqref{eqn:sce}.
To treat the full equation in the framework of quantum fields on a curved space-time can be a difficult
endeavour~\cite{Birrell:1982,Wald:1994}. But in the linearised theory of gravity~\cite{Misner:1973},
where the space-time metric is written as
\begin{equation}\label{eqn:lingrav}
 g_{\mu \nu} = \eta_{\mu \nu} + h_{\mu \nu} \,,
\end{equation}
the expansion in $h_{\mu \nu}$ is well-known to yield the gravitational wave equations at leading order.
This remains right for the semi-classical equation \eqref{eqn:sce} where one obtains~\cite{Misner:1973}
\begin{equation}
\label{eqn:grav-wave}
 \Box h_{\mu \nu} = - \frac{16 \pi G}{c^4}  \left(\bra{\Psi} \hat{T}_{\mu \nu} \ket{\Psi}
- \frac{1}{2} \eta_{\mu \nu} \bra{\Psi} \eta^{\rho \sigma} \hat{T}_{\rho \sigma} \ket{\Psi} \right) \,,
\end{equation}
imposing the de Donder gauge-condition $\partial^\mu(h_{\mu\nu}-\frac{1}{2}\eta_{\mu\nu} \eta^{\rho \sigma} h_{\rho \sigma}) = 0$.
Here $\Box$ denotes the d'Alembert operator.
Note that the energy-momentum tensor at this order of the linear approximation is that for flat space-time, while in~\eqref{eqn:sce}
it was still in curved space-time.
In the Newtonian limit, where $\bra{\Psi} \hat{T}_{00} \ket{\Psi}$ is large compared to the other nine
components of the energy-momentum tensor, equation \eqref{eqn:grav-wave} becomes the Poisson equation
\begin{equation}
\label{eqn:poisson}
 \nabla^2 V = \frac{4 \pi G}{c^2} \, \bra{\Psi} \hat{T}_{00} \ket{\Psi} 
\end{equation}
for the potential $V = -\frac{c^2}{2} h_{00}$. This is simply the usual behaviour of General Relativity
in the Newtonian limit: space-time curvature becomes a Newtonian potential sourced by the energy-density term
of the energy-momentum tensor.

This potential term now contributes to the Hamiltonian of the matter fields, which in turn yields the dynamics
in the \schr\ equation.
To be more specific, in the linearised theory of gravity \eqref{eqn:lingrav}, the interaction between gravity
and matter is given by the Hamiltonian~\cite{Maggiore:2008}
\begin{equation}
\label{eq:H_int_classical}
 H_\text{int} = -\frac{1}{2} \int \D^3 r \, h_{\mu \nu} \,{T}^{\mu \nu} \,.
\end{equation}
The quantisation of the matter fields then provides us with the corresponding operator:
\begin{equation}
\label{eq:H_int}
 \hat{H}_\text{int} = -\frac{1}{2} \int \D^3 r \, h_{\mu \nu} \,\hat{T}^{\mu \nu} \,.
\end{equation}
It is important to point out the difference to a quantised theory of gravity. In the latter,
$h_{\mu\nu}$ becomes an operator as well, simply by applying the correspondence principle to the
perturbation $h_{\mu\nu}$ of the metric---and thereby treating the classical $h_{\mu\nu}$ like a field living
\emph{on} flat space-time rather than a property \emph{of} space-time.
In contrast to this, $h_{\mu\nu}$ here remains fundamentally classical.
It is determined by the wave equations \eqref{eqn:grav-wave},
which are meant as classical equations of motion.

In the Newtonian limit, where $\hat{T}_{00}$ is the dominant term of the energy-momentum tensor, the
interaction Hamiltonian then becomes
\begin{equation}
\label{eqn:hint}
 \hat{H}_\text{int} = \int \D^3 r \, V \,\hat{T}^{00} = -G \int \D^3 r \, \D^3 r' \frac{\bra{\Psi}
\hat{\varrho}(\vec r')\ket{\Psi}}{\abs{\vec r - \vec r'}} \, \hat{\varrho}(\vec r) \,,
\end{equation}
where we already integrated equation \eqref{eqn:poisson} and used $\hat{T}_{00} = c^2 \hat{\varrho}$
in the non-relativistic limit.
The mass density operator $\hat{\varrho}$ is simply $m \hat{\psi}^\dagger \hat{\psi}$ when only one kind
of particle is present.
Therefore, following the standard procedure~\cite{Robertson:1972} we end up with the \sne\ in Fock space:
\begin{equation} \label{eqn:fock}
\begin{split}
\mathrm{i} \hbar \partial_t \ket{\Psi}
&= \Bigg[ \int \D^3 r \, \hat{\psi}^\dagger(\vec r) \left(-\frac{\hbar^2}{2m} \nabla^2 \right) \hat{\psi}(\vec r) \\
&\bleq - G\, m^2 \int \D^3 r  \, \D^3 r' \frac{\bra{\Psi}
\hat{\psi}^\dagger(\vec r') \hat{\psi}(\vec r')\ket{\Psi}}{\abs{\vec r - \vec r'}} \, \hat{\psi}^\dagger(\vec r) \hat{\psi}(\vec r) \Bigg] \ket{\Psi}\,.
\end{split}
\end{equation}

In the non-relativistic limit the number of particles is conserved and we can without further assumptions
go over to the first-quantised form. For an $N$-particle state
\begin{equation} \label{eqn:n-particle-state}
\ket{\Psi_N} = \frac{1}{\sqrt{N!}} \left\{\int\prod_{i=1}^N  \D^3 r_i \right\} \, \Psi_N(t,\vec r_1, \dots, \vec r_N)
\hat{\psi}^\dagger(\vec r_1) \cdots \hat{\psi}^\dagger(\vec r_N) \ket{0} \,,
\end{equation}
where $\Psi_N(t,\vec r_1, \dots, \vec r_N)$ is the $N$-particle wave-function, the expectation value is
\begin{equation}
\bra{\Psi_N} \hat{\psi}^\dagger(\vec r) \hat{\psi}(\vec r)\ket{\Psi_N} = \sum_{j=1}^N
\left\{\int\prod_{i=1 \atop i \not= j}^N  \D^3 r_i \right\}
\abs{\Psi_N(t,\vec r_1, \dots , \vec r_{j-1} , \vec r, \vec r_{j+1} \dots , \vec r_N)}^2 \,.
\end{equation}
Equation~\eqref{eqn:fock} with the state \eqref{eqn:n-particle-state} inserted therefore yields
the $N$-particle \sne~\cite{Diosi:1984}
\begin{equation}
\label{eq:n-particleSN}
\begin{split}
&\mathrm{i}\hbar\partial_t\Psi_N(t;\vec r_1,\cdots,\vec r_N)
=\biggl(-\sum_{i=1}^N\frac{\hbar^2}{2m}\nabla^2_i \biggr.\\
\biggl.&-G m^2 \sum_{i=1}^N\sum_{j=1}^N \left\{\int\prod_{k=1}^N  \D^3 r'_k\right\}
 \frac{\abs{\Psi_N(t;\vec r_1',\cdots,\vec r_N')}^2}{\abs{\vec r_i - \vec r'_j}}
\biggr)\Psi_N(t;\vec r_1,\cdots,\vec r_N)\,,
\end{split}
\end{equation}
and in the one-particle case the \sne~\eqref{eqn:sn} follows immediately.

We therefore unavoidably
obtain the \sne\ for non-relativistic quantum matter \emph{if} the initial assumptions are correct: that gravity
is fundamentally classical, and that the semi-classical Einstein equations \eqref{eqn:sce} describe its coupling
to matter.
In this precise sense, the \sne\ \emph{does} follow from fundamental principles.
Whether or not these principles and the underlying assumptions are correct, is a different story which,
eventually, will be decided by experiments.

\subsection{The \sne\ as a Hartree approximation} \label{sec:hartree}
It is important to stress that the \sne\ also appears in a different context, however with a totally different meaning.
Assuming that gravity is fundamentally described by a quantum theory in which the metric perturbation
$h_{\mu \nu}$ turns into a linear operator~\cite{Donoghue:1994}, then very similar to what we have
in equation~\eqref{eqn:grav-wave}, in the weak-field limit one gets the linearised equation:
\begin{equation}
\label{eqn:QM_grav-wave}
\Box \hat{h}_{\mu \nu} = - \frac{16 \pi G}{c^4} \left(\hat{T}_{\mu \nu}
- \frac{1}{2} \eta_{\mu \nu} \eta^{\rho \sigma} \hat{T}_{\rho \sigma}  \right) \,,
\end{equation}
where $\hat{h}_{\mu \nu}$ now is a linear quantum operator. 
In complete analogy to equation \eqref{eqn:poisson}, one obtains the potential $\hat{V}$ in the Newtonian limit where:
\begin{equation}
 \nabla^2 \hat{V} = \frac{4 \pi G}{c^2} \, \hat{T}_{00}. 
\end{equation}
Note that, contrary to equation \eqref{eqn:poisson}, $\hat{V}$ also carries a hat now, i.\,e. it is considered a quantum operator.
Then, the corresponding interaction Hamiltonian reads as:
\begin{equation}
\label{eqn:hint_hartree}
 \hat{H}_\text{int} = \int \D^3 r \, \hat{V} \,\hat{T}^{00} = -G \int \D^3 r \, \D^3 r' \,\frac{\hat{\varrho}(\vec r')\, \hat{\varrho}(\vec r)}{\abs{\vec r - \vec r'}} \,.
\end{equation}
Note that
the interaction term in equation \eqref{eqn:hint_hartree} is in the second-quantised formalism, and we can
again write down the full \schr\ equation in Fock space:
\begin{equation} \label{eqn:fock-hartree}
\begin{split}
\mathrm{i} \hbar \partial_t \ket{\Psi}
&= \Bigg[ \int \D^3 r \, \hat{\psi}^\dagger(\vec r) \left(-\frac{\hbar^2}{2m} \nabla^2 \right) \hat{\psi}(\vec r) \\
&\bleq - G\, m^2 \int \D^3 r  \, \D^3 r'
\frac{\hat{\psi}^\dagger(\vec r') \hat{\psi}(\vec r') \hat{\psi}^\dagger(\vec r) \hat{\psi}(\vec r)}%
{\abs{\vec r - \vec r'}} \Bigg] \ket{\Psi}\,.
\end{split}
\end{equation}

The corresponding $N$-body Schr\"odinger equation in first-quantised formalism is given by
\begin{subequations}
\begin{equation}
\label{eqn:sch_N}
\mathrm{i}\hbar\frac{\partial}{\partial t}\Psi_N(t,\vec r_1,\dots,\vec r_N)
=\hat{H}_N\,\Psi_N(t,\vec r_1,\dots,\vec r_N) \,,
\end{equation}
where
\begin{equation}
\hat{H}_N=-\sum_{j=1}^N \frac{\hbar^2}{2 m_j} \nabla^2_j
- G m^2 \, \sum_{i \not= j} \frac{1}{\abs{\hat{\vec r}_i - \hat{\vec r}_j}} \,.
\end{equation}
\end{subequations}
The last term is the contribution of $\hat{H}_\text{int}$ given in equation \eqref{eqn:hint_hartree}.
In the last term on the right-hand side of the above equation, the infinite terms with $i=j$ are treated by standard
renormalisation and regularisation techniques at the second-quantised level and they appear as
a mass renormalisation.

Then, in the case of many-particle systems for $N\to\infty$, and assuming that all particles have the same mass $m$,
one can obtain the nonlinear Hartree equation as the mean-field limit of equation \eqref{eqn:sch_N}:
\begin{equation}
\label{eqn:hartree}
\mathrm{i}\hbar \frac{\partial}{\partial t}\psi(t,\vec{r})=
\left(-\frac{\hbar^2}{2m}\,\nabla^2
-Gm^2\int \,\D^3 r'\,\frac{\,\abs{\psi(t,\vec{r}')}^2}{\abs{\vec r - \vec r '}}
\right)\psi(t,{\vec r}) \,.
\end{equation}
Note that the precise mathematical derivation is not a trivial endeavour and involves implementing the quantum
Bogoliubov--Born--Green--Kirkwood--Yvon (\abbrv{BBGKY}) hierarchy in the limit
$N\to\infty$~\cite{Erdoes:2001,Bardos:2002,Benedikter:2013}.

Formally, equation~\eqref{eqn:hartree} is the same as equation~\eqref{eqn:sn}. 
However, one should keep in mind that equation~\eqref{eqn:hartree} is derived as the mean-field limit of an $N$-body
linear \schr\ equation and it is only an {\it effective} description to the zero-th order of the dynamics of the
$N$-body system. At the level of the full $N$-body system the dynamics are still linear, as given in
equation~\eqref{eqn:sch_N}.
The centre-of-mass wave-function is described by a \emph{free} \schr\ equation.

In contrast to this, in semi-classical gravity we have nonlinearity even at the level of the $N$-body dynamics,
yielding contributions to the centre-of-mass motion \cite{Giulini:2014}.
Therefore, while in the relative motion the non-gravitational interactions dominate the dynamics,
in the centre-of-mass dynamics gravitation is the only interaction and can therefore lead to observable effects.

\subsection{The \sne\ and the meaning of the wave-function}

Considering the expectation value of the energy-momentum tensor as the source of the gravitational field,
the wave-function in the \sne\ coincides with the gravitational
mass distribution, whose different parts
attract gravitationally. Following this line of reasoning, a particle would not be really point-like. It would be a
wave packet, which tends to spread out in space, according to the usual dynamical term in the \schr\ equation,
and shrink, according to the \schr--Newton term. An equilibrium is reached when the two effects compensate.
This equilibrium state is a soliton, which is physically perceived as a particle.

It can be argued~\cite{Adler:2007,Anastopoulos:2014}
that this mass-density interpretation of the wave-function is in contradiction with the usual probabilistic 
interpretation, according to which $\abs{\psi(\vec r)}^2$ represents the probability density of finding the
(point-like) particle in $\vec r$, at the end of a position measurement.
On this ground, one could tend to dismiss the \sne. In fact, \schr\ himself, who had a quite similar interpretation
of the wave-function in mind, already noted~\cite{Schroedinger:1927} that in this picture a self-interaction of
the wave-function seems to be a natural consequence for the equations to be consistent from a field-theoretic point
of view. But he also noticed that, for reasons he could not understand, in the case of electrodynamical interactions
such nonlinearities in the \schr\ equation would give totally wrong numbers for known phenomenology, e.\,g. the
hydrogen spectrum. However, in the case of gravity such an argument does not hold, since all effects of the \sne\ are
far below what has been experimentally observed to date.

It is clear that in a theory based on the \sne, the wave-function must bear both roles:
of suitably describing the mass density, \emph{and} of providing the probability distribution of outcomes of
measurements. The question arises: can this be consistently achieved? A positive answer is provided by
collapse models; cf.~\cite{Ghirardi:1986,Ghirardi:1990a,Adler:2007b,Adler:2009,Bassi:2003,Bassi:2013}
and section~\ref{sec:comparison}. In these models, the primary role of the wave-function is to describe matter,
meaning with it that a particle is not a point-like particle but is no more and no less than what the wave-function
says: a wave packet, which tends to be localised thanks to the collapse mechanism.
The important point is that when a particle wave packet interacts with a device which measures its position,
the collapse dynamics will say that at the end of the measurement the outcomes will be
distributed randomly, according to the Born rule~\cite{Bassi:2007}.
Therefore, the Born rule is not an additional postulate which assigns a probabilistic role to the
wave-function. It is a by-product of the dynamics, when applied to what we typically refer to as
measurement processes. It is a handy way---and nothing more---to directly calculate the probability of
the outcomes of measurements, rather than solving each time the full equations of motion.

Eventually, the same situation should occur for the \sne, when incorporated in a fully consistent theory.
In fact, at the end of the day, the goal of this type of research is to \emph{explain} the collapse of the
wave-function (and with it the Born rule) from an underlying physical principle, which in this case is gravity.

\section{Wave-function collapse and the \sne}\label{sec:collapse}
\begin{figure}
\hspace{2.5cm}{\includegraphics[width=10cm,keepaspectratio=true]{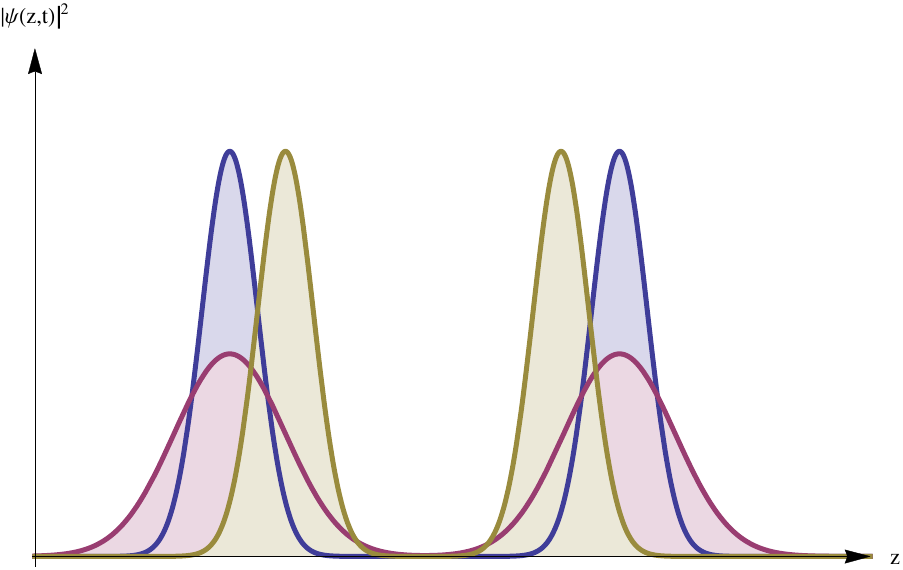}}
\caption{Given a state which is in a superposition of two wave packets (blue lines) we compare the normal
\schr\ evolution (purple lines) with the one given by the \sne\ (yellow line). The \schr--Newton dynamics
imply an attraction between the different parts of the wave-function which result in both a self-focusing
effect on each of the wave packets and an attraction between the two wave packets.}
\label{one}
\end{figure}

It has been claimed that the \sne\ provides an explanation for the collapse of the
wave-function~\cite{Diosi:1984,Penrose:1996,Penrose:1998,Penrose:2014}.
In this section we will elaborate on this. 
The nonlinear gravitational interaction implies an attraction among different
parts of the wave-function. When the wave-function of the system is given by a single wave packet, this effect
amounts to an inhibition of the free-spreading of the wave packet, thus resulting in a self-focusing (or say,
shrinking) of it for sufficiently high masses~\cite{Carlip:2008,Harrison:2003}.
Additionally, in a system which has been prepared in a spatial superposition of two wave packets at different
locations, the nonlinear interaction also implies an attraction between those wave packets
(see figure \ref{one}). 

The strength of the nonlinearity in the \sne\ depends on the size of the system, in particular on its mass.\footnote{%
Denoting the size of a (homogeneous, spherical) particle by $R$, the width of the wave packet by $\sigma$,
its mass by $m$, and by $l_p$ and $m_p$ the Planck length and Planck mass, respectively, one finds that
in the case $R \ll \sigma$ significant deviations from linear \schr\ dynamics occur if
$m^3 \, \sigma  \gtrsim  m_p^3 \, l_p$, and in the case $R \gg \sigma$ they occur if
$m^3 \, \sigma^2 / R  \gtrsim  m_p^3 \, l_p$~\cite{Giulini:2011,Giulini:2014}.
}
Now we comment on why the attraction between different parts of the wave-function does not
account for the usual collapse of the wave-function.
Accordingly, exactly as in standard Quantum Mechanics, the collapse postulate, including the Born rule,
must be supplemented to the \sne\ in order to provide a full description of experimental situations.

Let us consider an experiment where a particle's position is measured. Take an initial superposition state for
the particle
\begin{equation}
 \psi(\vec r)=\frac{1}{\sqrt{2}}(\psi_1(\vec r)+\psi_2(\vec r)) \,,
\end{equation}
where $\psi_1(\vec r)$ and $\psi_2(\vec r)$ are wave packets well localised around $\vec r_1$ and $\vec r_2$,
respectively. 
During the measurement, this state couples with the massive measuring instrument (say, a pointer) as follows:
\begin{equation}\label{eqn:spatial-superpos-pointer}
 \Psi(\vec r,\vec R)=\frac{1}{\sqrt{2}}(\psi_1(\vec r)\Phi_1(\vec R)+\psi_2(\vec r)\Phi_2(\vec R)) \,,
\end{equation}
where $\Phi_{1}(\vec R)$ and $\Phi_{2}(\vec R)$ are two localised wave-functions of the pointer, centred
around $\vec R_1$ and $\vec R_2$, respectively.\footnote{In practice these wave-functions will have a finite
overlap, but this can be neglected for the purpose of the argument we provide here.}
The positions $\vec R=\vec R_{1,2}$
correspond to the particle being around positions $\vec r_{1,2}$.
Since the pointer is a classical system, according to the orthodox interpretation, the wave-function collapses
at $\vec R=\vec R_{1}$ or $\vec R=\vec R_2$, revealing in this way the outcome of the measurement. This
means that the particle is found half of the times around the position $\vec r_1$ and half of the times around
the position $\vec r_2$. 

According to the \sne\ without the standard collapse postulate, on the other hand,
a superposition state as in \eqref{eqn:spatial-superpos-pointer} implies a gravitational attraction also
between the spatial wave packets $\Phi_{1}$ and $\Phi_2$ representing the massive pointer. The wave-function of
the pointer would always ``collapse'' to the average position $(\vec R_1 + \vec R_2)/{2}$, simply due to the
symmetry of the deterministic dynamics and the initial state.
Numerical simulations confirm this behaviour of spatial superpositions collapsing to an average 
position~\cite{Harrison:2003}. Such a behaviour is however in obvious
contradiction with the standard collapse postulate, as well as with our everyday experience, where the pointer
is found with equal probability either at
$\vec R=\vec R_1$ or at $\vec R=\vec R_2$, and never in the middle.

Moreover the \sne\ is deterministic and as such it cannot explain why quantum measurements occur randomly,
distributed according to the Born rule. Therefore, the \sne\
explains neither the standard collapse postulate nor the Born rule;
one still needs both to describe experimental results, as long as no additional collapse prescription
is added.

\section{Superluminal effects in the \sne} \label{sec:thought-experiment}

As we pointed out in the introduction, semi-classical gravity together with the standard collapse postulate
leads to violation of local energy-momentum conservation. But even if we take the \sne\ as a hypothesis,
without relating it to semi-classical gravity, together with the standard collapse
postulate, it leads to superluminal effects, as all
nonlinear deterministic \schr\ equations do~\cite{Gisin:1989}.
In this section, we discuss a concrete thought experiment, that shows how
the \sne\ implies faster-than-light signalling.

Consider a spin 1/2 particle in a Stern--Gerlach apparatus with a magnetic force in the $z$-direction, where the
position of the particle along the $z$-axis is finally observed at the detector. We denote the spin eigenstates
along the $z$-axis by $|z_\pm\rangle$ and along the $x$-axis by
$|x_\pm\rangle=\frac{1}{\sqrt{2}}(|z_+\rangle\pm|z_-\rangle)$. We assume that there is no coupling
of the spin to
the spatial wave-function due to gravity, implying that the spatial wave-function evolves according to
the \sne\ while the spin wave-function evolves according to a linear \schr\ equation.

\begin{figure}
\centering
\subfloat[Particle with well defined spin along $z$]{%
\includegraphics[width=65mm,keepaspectratio=true]{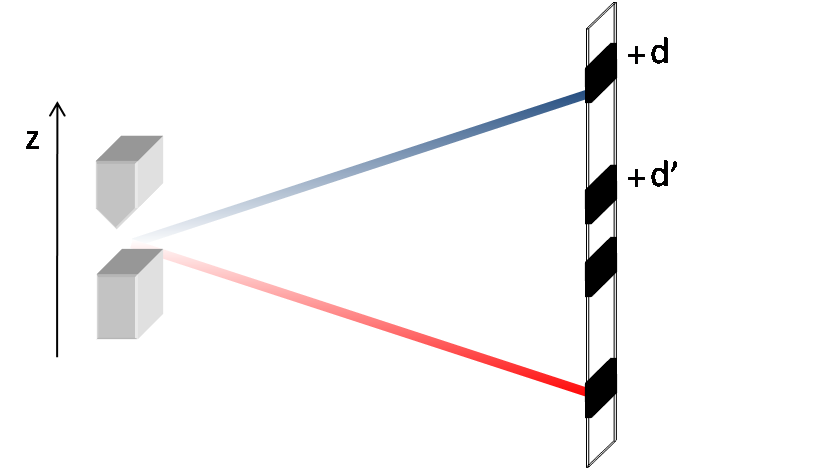}}
\hspace{1cm}
\subfloat[Superposition of states with well defined spin along $z$]{%
\includegraphics[width=65mm,keepaspectratio=true]{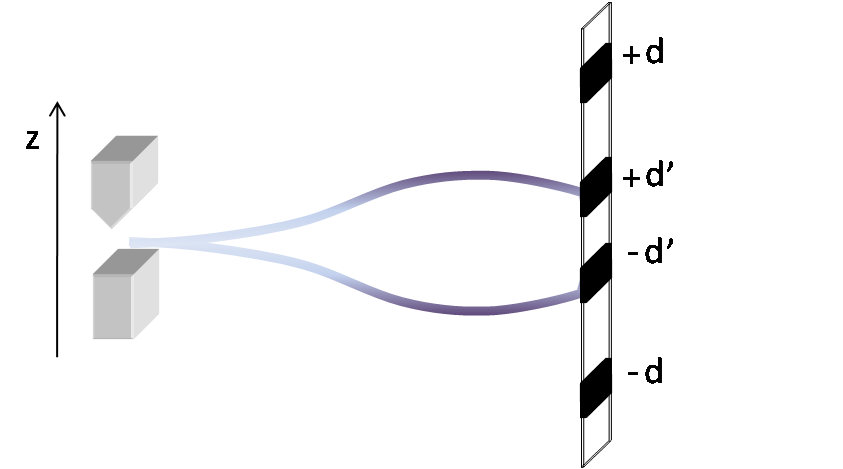}}
\caption{Predictions of the \sne\ for a particle passing through a Stern--Gerlach apparatus. While the state with
well defined spin on the left is found either around the position $+d$ or $-d$, the superposition on the right
is found either around the position $+d'$ or $-d'$, because of the attraction among different parts of the
wave-function due to the nonlinear interaction.}
\label{two}
\end{figure}

For the initial spin state $|z_\pm\rangle$ the state after the particle passing through the
Stern--Gerlach apparatus is
$\Psi(t,\vec r)=\psi_\pm(t,\vec r)\otimes\ket{z_\pm}$.
The spatial state evolves as
\begin{equation}
\label{eq:z}
\mathrm{i}\hbar\frac{\partial}{\partial t}\psi_\pm(t,\vec r)=
\left(-\frac{\hbar^2}{2m}\nabla^2
-Gm^2\int \,\D z' \,\frac{|\psi_\pm(t,\vec r')|^2}{|\vec r-\vec r'|}
\right)\psi_\pm(t,\vec r) \,,
\end{equation}
with the initial conditions $\psi_\pm(0,\vec r)=e^{\pm ik_zz}\phi(\vec r)$,
where $k_z$ is the momentum induced by the Stern--Gerlach field and $\phi$ is the initial wave packet.
Without loss of generality we assume that $\phi$ is a stationary solution of the \sne\ and therefore
the evolution before passing the apparatus plays no role. After passing, the nonlinear gravitational
interaction leads to a self-focusing that inhibits the free spreading of the wave packet. The centre of mass
moves upwards if the initial state is $\psi_+$ or downwards if it is $\psi_-$. The wave packets
$\psi_\pm(t,z)$ are finally observed at the detector positions $\pm d$ (see figure \ref{two}a).

Now consider the opposite case with initial states $|x_\pm\rangle$. Then the state of the particle after
passing through the Stern--Gerlach apparatus is
\begin{equation}
\Psi(t,\vec r)=\frac{1}{\sqrt 2}\left(\chi_+(t,\vec r)\otimes\ket{z_+}+\chi_-(t,\vec r)\otimes\ket{z_-}\right) \,.
\end{equation}
In this case, the time evolution for the states $\chi_+$ and $\chi_-$ as predicted by the \sne\ is given by
\begin{multline}\label{eqn:psiplusminus}
\mathrm{i}\hbar\frac{\partial}{\partial t}\chi_\pm(t,\vec r)
=
\Bigg(
-\frac{\hbar^2}{2m}\nabla^2
-\frac{Gm^2}{2}\int \D^3 r' \,\frac{|\chi_\pm(t,\vec r')|^2}{|\vec r-\vec r'|}\\
-\frac{Gm^2}{2}\int \D^3 r' \,\frac{|\chi_\mp(t,\vec r')|^2}{|\vec r-\vec r'|}
\Bigg)\chi_\pm(t,\vec r),
\end{multline}
with initial conditions $\chi_\pm(0,\vec r)=e^{\pm ik_zz}\phi(\vec r)$ as before.  
The second term on the right-hand side of equations \eqref{eqn:psiplusminus} is the self-focusing force,
while the third term corresponds to the attraction between the two wave packets $\chi_+$ and $\chi_-$.
Due to this attraction, the wave packets are finally observed at the detector positions $\pm d'$,
where $d'<d$ (see figure \ref{two}b).
Accordingly, the \sne\ gives different predictions, compared to the linear Schr\"odinger equation,
for a Stern--Gerlach $z$-spin measurement with initial states $|x_\pm\rangle$, while it coincides with the
linear Schr\"odinger equation for initial states $|z_\pm\rangle$ if one neglects the change in the width of
wave packets for the time-scale of the experiment.

The aforementioned gravitational attraction can, at least in principle, be exploited experimentally to
distinguish standard quantum theory from the \sne.
For this purpose, we provide a closer look at the conceptual implications of this effect in connection with
entanglement and the non-local nature of the quantum measurement process. 
This effect, that makes use of the nonlinear \schr--Newton dynamics, can then, in principle, be used in order
to send signals faster than light.

\begin{figure}
\centering
\includegraphics[width=13cm,keepaspectratio=true]{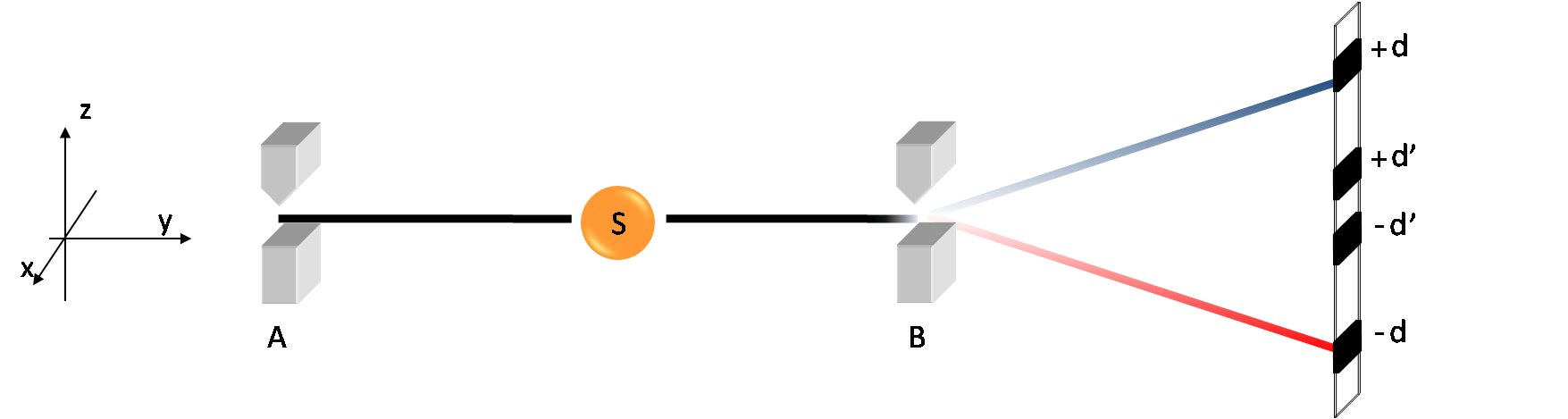}
\includegraphics[width=13cm,keepaspectratio=true]{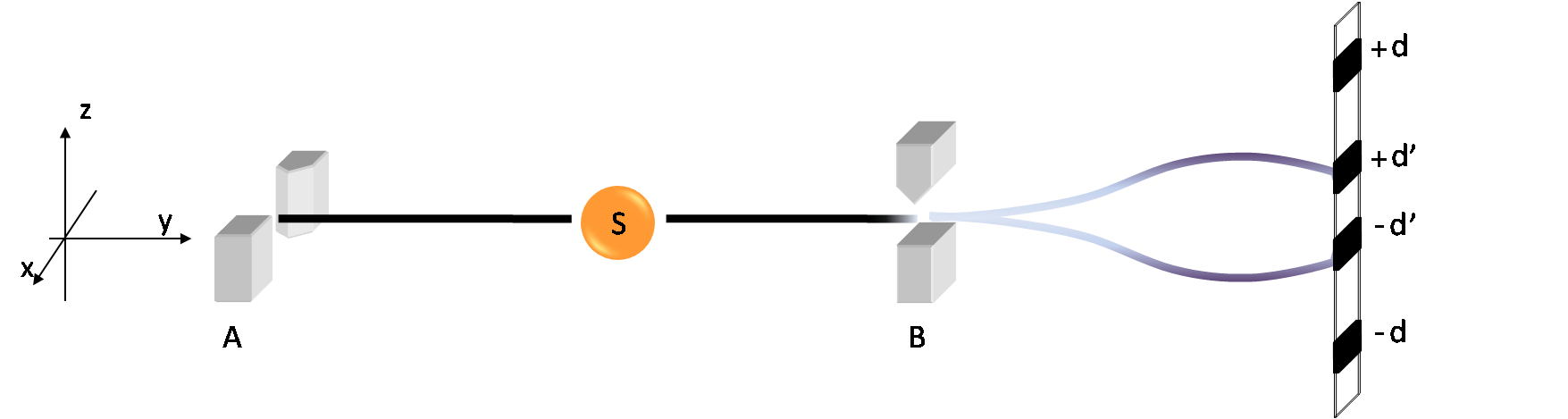}

\caption{Two entangled particles are sent from the source (S) to Alice (A) and Bob (B). Alice is the first one
who performs the spin measurement, which can be done along the $z$ or $x$ direction. In the first case,
the particle on Bob's side has a well defined spin along the $z$ direction, and Bob will find the particle
around either points $\pm d$. On the contrary, in the second case the particle on Bob's side is in a
superposition of states with well defined spin along the $z$, which implies that Bob will find the particle
around either points $\pm d'$. Since the distance between Alice and Bob can be arbitrary large, this
setup implies faster-than-light signalling.\label{three}}
\end{figure}

For that purpose consider a typical EPR set-up, where two particles move in opposite directions toward
two Stern--Gerlach devices (see  figure \ref{three}). They are initially prepared
in a singlet spin state:
\begin{eqnarray}
\Psi(t=0,\vec r_A,\vec r_B)=\frac{1}{\sqrt{2}}\left(
|z_+\rangle_A\otimes|z_-\rangle_B-|z_-\rangle_A\otimes|z_+\rangle_B
\right)\otimes\phi_1(\vec r_A)\otimes\phi_2(\vec r_B) \,,
\end{eqnarray} 
Consider an observer (Alice) on the left-hand side of our experimental
Stern--Gerlach setting. Alice performs spin measurements either in $z$- or in $x$-direction, always before the
other entangled particle enters the Stern--Gerlach apparatus on the opposite side, where a second observer (Bob)
also is making spin measurements. According to the discussion of the previous section, the measurement
is described by the standard collapse rule. Bob always measures the spin in the $z$-direction.
If Alice measures the spin in the $z$-direction, then Bob's particle is prepared in one of the states $|z_\pm\rangle$.
Therefore, Bob will detect the particle at positions $\pm d$ (see figure \ref{three}).
On the other hand, when Alice measures the spin in the $x$-direction, Bob's particle will be prepared in the state
$|x_\pm\rangle$, meaning that Bob will detect the particles at positions $\pm d'$ (see figure \ref{three}).
Although, in general, due to the weakness of the gravitational interaction, the difference between $d$ and $d'$ will be 
incredibly hard to measure,\footnote{%
%
%
To give some numbers, consider a particle travelling with velocity $v \ll c$ over distance $S$ to the Stern--Gerlach
device, where it is split in a superposition with spatial separation $d_0$. The two states in the superposition then
travel parallel to each other over a distance $s$. Considering the Newtonian gravitational acceleration one obtains
$\Delta d \approx G m s^2/(2 v^2 d_0^2)$ as an estimate for the distance they are shifted.
For the communication to be superluminal the time it would take for a photon to reach the Stern--Gerlach
device, $S/c$, must be larger than than the time $s/v$ it takes for the particle to acquire a measurable displacement
$\Delta d$. Therefore $S > c \, d_0 \, \sqrt{2 \Delta d/(G m)} $ must hold.
Current experiments can achieve $m \approx 10\,000$\,\atomicmass\ and
$d_0 \approx \Delta d \approx 1$\,\micro\meter, yielding a minimum distance of about one light-year for
superluminal signalling with state-of-the-art technology.}
%
%
in principle, Bob can identify which measurement Alice performed by only looking at the position where
he observes the particle hitting the screen. Therefore the \sne\ allows to send signals faster than the speed
of light.

An important remark is at order. The kind of faster-than-light signalling discussed in this section is an effect
of the instantaneous collapse of the wave-function (as a result of Alice's measurement), together with the nonlinear
character of the dynamics described by the \sne. Therefore, even if one describes the whole situation in a fully
relativistic way (i.\,e. by some sort of ``Dirac--Newton equation'', which one could eventually obtain by
applying \eqref{eqn:sce} to a Dirac field), one would not get rid of the instantaneous collapse of the wave-function
upon measurement, nor of the nonlinear character of the dynamics. What would change is the way the two parts of
the superposition attract each other: in the \sne\ this attraction is instantaneous, while in the relativistic
framework it would likely have a finite speed. This amounts in slight differences in the self-gravitation
effects, which do not play any important role for the argument proposed here. As long as there is some measurable
effect of self-gravitation, Bob can always exploit it to figure out Alice's measurement setting, and thereby
receive a signal with the ``speed of collapse'' (which is infinite in the standard collapse prescription and
has been shown to exceed the speed of light by orders of magnitude in a multitude of 
experiments~\cite{Stefanov:2002,Scarani:2014,Bancal:2012}).

Contrary to this situation, it has been widely studied how to modify \schr\ equation by adding nonlinear and
stochastic terms, in order to describe the collapse of the wave-function, while avoiding superluminality.
Collapse models provide a mathematically consistent phenomenology of this
type~\cite{Ghirardi:1986,Ghirardi:1990a,Adler:2007b,Adler:2009,Bassi:2003,Bassi:2013}. In the next section,
we elaborate on this issue and discuss the connection of the \sne\ with collapse models.

\section{Comparison between the \sne\ and collapse equations} \label{sec:comparison}

The reason why no superluminal effects appear in collapse
models~\cite{Ghirardi:1986,Ghirardi:1990a,Adler:2007b,Adler:2009,Bassi:2003,Bassi:2013}
is that the nonlinear modification is balanced by appropriate stochastic contributions.
At the density-matrix level the collapse dynamics
are generally of Lindblad type:
\begin{equation} \label{eq:lin}
\frac{\partial}{\partial t}\hat{\rho}_t  =  -\frac{\mathrm{i}}{\hbar}[\hat{H},\hat{\rho}_t] 
+ \gamma \int\,\D^3 k\left(\hat{L}({\vec k})\hat{\rho}_t\hat{L}({\vec k})^{\dagger}
- \frac12 \hat{L}({\vec k})^{\dagger}\hat{L}({\vec k})\hat{\rho}_t
- \frac12 \hat{\rho}_t\hat{L}({\vec k})^{\dagger}\hat{L}({\vec k})\right)\,,
\end{equation}
where both the \emph{linear} operator $\hat{L}$ and the \emph{real} coupling constant $\gamma$ can be chosen
arbitrarily and must be specified in concrete models.
As one cause, at the density-matrix level, stochastic terms perfectly cancel all nonlinear terms.

At the wave-function level, on the other hand, both nonlinear and stochastic contributions appear,
and the dynamical equation is of the following type:
\begin{eqnarray}
\label{eq:wave}
\frac{\partial}{\partial t}|\psi_t\rangle &=&
\left[
-\frac{\mathrm{i}}{\hbar}\hat{H}
+\sqrt{\gamma}\int \D^3 k\,\left(\hat{L}({\vec k})-\ell_t({\vec k})
\right)\xi(t,\vec k)\right.\\\nonumber&&\left.\quad
-\frac{\gamma}{2}\int \D^3 k\,\left(
\hat{L}(\vec k)^\dagger\hat{L}(\vec k)-2\ell_t(\vec k)\hat{L}(\vec k)
+|\ell_t(\vec k)|^2\right)
\right]|\psi_t\rangle \,.
\end{eqnarray}
In this modified \schr\ equation, nonlinearities are introduced by the expectation-value term
\begin{equation}
\ell_t(\vec k)  \equiv  
\frac{1}{2}\left(\langle\hat{L}^{\dagger}(\vec k)\rangle_t
+\langle\hat{L}(\vec k)\rangle_t\right);\qquad
\qquad \text{with}\qquad \langle\hat{L}(\vec k)\rangle_t=\langle\psi_t|\hat{L}(\vec k)|\psi_t\rangle \,
\end{equation}
coupled to a random noise $\xi(t,\vec k)=\D W(t,\vec k)/\D t$, where $W(t,\vec k)$ are independent
Wiener processes.

How does this compare to the \sne? In order to see this, first note that the Fourier transform of the
Newtonian gravitational potential term is given by~\cite{Jackson:1962}
\begin{equation}
 -\frac{G\, m^2}{\abs{\vec r - \vec r'}} = -\frac{G\, m^2}{2 \pi^2} \int \D^3 k \,
\frac{\exp\left(\mathrm{i} \, \vec k \cdot (\vec r - \vec r')\right)}{k^2} \,.
\end{equation}
If we now introduce the linear operator
\begin{equation}
\label{eq:lin_operator}
\hat{L}({\vec k})=m\,\frac{\exp\left(\mathrm{i} \, \vec k \cdot \hat{\vec r}\right)}{k} \,,
\end{equation}
we can write the \sne\ in terms of solely this operator and its expectation value:
\begin{equation}
\label{eq:sn_operator}
\frac{\partial}{\partial t}|\psi_t\rangle=
\left(-\frac{\mathrm{i}}{\hbar}\hat{H}+\mathrm{i}\frac{G}{2\pi^2\hbar}\int\,
\D^3 k\,\langle\hat{L}^\dagger({\vec k})\rangle_t\,\hat{L}({\vec k})\right)|\psi_t\rangle \,.
\end{equation}
As we see, the equation is of a completely different structure with respect to the collapse
equation~\eqref{eq:wave}. In particular, the coupling constant in front of the nonlinear
term is imaginary while $\gamma$ in equation~\eqref{eq:wave} is real.

We could obtain a term as in equation \eqref{eq:sn_operator}, if in equation~\eqref{eq:wave}
we replace the noise field by
\begin{equation}
\xi(t,\vec k)\rightarrow\xi'(t,\vec k)+\mathrm{i}\langle\hat{L}^\dagger(\vec k)\rangle_t \,,
\end{equation}
thereby introducing an imaginary drift. The \schr--Newton term would however appear with
many other terms which completely change the dynamics.

The collapse equation that comes closest to combining equation~\eqref{eq:wave} with gravity
is the collapse model by Di{\'o}si~\cite{Diosi:1987,Diosi:1989,Diosi:2007}. This model, in its original
form, is obtained by using the operator \eqref{eq:lin_operator} and the constant
$\gamma = G/(2 \pi^2 \hbar)$ in equation~\eqref{eq:wave}.

This equation, however, yields an infinite rate for the energy exchange
(i.\,e., $\mathrm{Tr}(\hat{H}\,\D\hat{\rho}_t/\D t)$ diverges). To avoid
this problem, one needs to modify the Lindblad operator, introducing a cut-off $R_0$:
\begin{equation}
\label{eq:lin_operator_cutoff}
\hat{L}(\vec k)=m\,\varrho(k,R_0)\,\frac{\exp\left(\mathrm{i}\,\vec k\cdot\hat{\vec r}\right)}{k} \,.
\end{equation}
A reasonable choice for the cut-off is $\varrho(k,R_0)\sim e^{-k^2R_0^2}$.
The value of $R_0$ was originally chosen equal to the
classical size of a nucleon ($\sim 10^{-15}\,$m)~\cite{Diosi:1987,Diosi:2007}.
Later on, it was shown~\cite{Ghirardi:1990}
that such a cut-off is too small, and still giving rise to an unacceptable energy increase of
$\sim 10^{-4}$\,\kelvin\per\second\ for a proton. This problem can be fixed by choosing a
much larger cut-off, e.\,g. $R_0\sim 10^{-7}$\,\meter, bringing the energy increase
down to about $10^{-28}$\kelvin\per\second.

Although such a model provides a consistent description of the wave-function collapse which does not
allow for faster-than-light signalling, its relation to gravity is restricted to the appearance of the
gravitational constant $G$ in the coupling. It is an effective model which is not derived from
known fundamental principles of physics and it is not clear how this could be done.
In fact, the dynamical equation~\eqref{eq:wave}
with the operator~\eqref{eq:lin_operator_cutoff} is simply postulated. In particular, the newly introduced free
parameter $R_0$ is not related to gravity, and the origin of the stochastic term $\xi$
remains unresolved.

\section{Conclusions}

The main goal of this paper was to straighten out the conceptual status of the \sne, and thereby
also clarifying the statements already made in~\cite{Anastopoulos:2014,Anastopoulos:2014a,Giulini:2012,Giulini:2014,Diosi:1984}.
We have shown that the \sne\ follows without further assumptions from a semi-classical theory of
gravity, i.\,e. a theory which treats gravity as fundamentally classical with quantum matter coupled to it via the
semi-classical Einstein equations \eqref{eqn:sce}.
The evolution according to the \sne\ differs from the
linear \schr\ equation in two respects. First, the \schr--Newton dynamics lead to a self-focusing
of wave packet solutions for the centre of mass, as it has been studied in \cite{Carlip:2008,Giulini:2011,Giulini:2013}.
Second, in contradiction to the probability interpretation of the wave-function, a spatial superposition
state will reveal a Newtonian gravitational attraction of different parts of the wave-function.

The \sne\ does, however, not serve the purpose it was originally
considered for, namely of providing an explanation for the wave-function collapse~\cite{Penrose:1998}.
A collapse prescription,
either in terms of the Copenhagen collapse postulate or in terms of an objective collapse model, is still
necessary to relate solutions of the \sne\ to outcomes of measurements.
Moreover, we explicitly have shown that with the conventional collapse prescription, the \schr--Newton
dynamics unavoidably lead
to the possibility of faster-than-light signalling.

A possible way to describe the collapse of the wave-function without having faster-than-light signalling
is given by collapse models.
These models avoid these superluminal
effects because of the random nature of the collapse dynamics. Therefore the question arises, if the
gravitational self-interaction of the wave-function according to the \sne\ can be brought together with the
collapse dynamics. But the way collapse models introduce nonlinearities is very different from that of
the \sne.
Also, the already known collapse models that are inspired by gravity~\cite{Diosi:1987,Diosi:1989,Diosi:2007}
do not fulfil the purpose of combining both ideas since they do not give a satisfactory explanation of how
the interactions that lead to the collapse derive from gravity.

Even if the Newtonian gravitational interaction could be consistently included in collapse models, the presence
of both \schr--Newton and collapse terms would not
resolve the main open question regarding collapse models, namely the physical nature of the stochastic
field causing the collapse. If one wants to attribute the collapse to gravity, the gravitational interaction has
to account for random effects, either instead of or on top of the semi-classical theory.

\section*{Acknowledgements}
We would like to thank the anonymous referees for helpful comments.
We gratefully acknowledge funding and support through the EU project \abbrv{NANOQUESTFIT},
the \abbrv{COST} Action MP1006 ``Fundamental Problems in Quantum Physics'',
the John Templeton foundation (grant 39530), and the  Istituto Nazionale di Fisica Nucleare (\abbrv{INFN}).

\appendix
\section*{\appendixname}

\subsection{Can gravity be fundamentally classical?}\label{app:semiclass-arguments}

The question of the validity of the \sne\ as a correction to the
linear one-particle \schr\ equation is directly related to the question if there is any need for a quantisation
of the gravitational field.
The great success of quantum theory led to the prevalent belief that such a quantum theory of gravity
must be found, by modifying General Relativity in order to make it compatible with linear quantum theory.
But there is no a priori argument rendering this approach more valid than the opposite one, i.\,e. retaining
the classical structure of gravity and modifying quantum theory. One should keep in mind that, after all,
the gravitational field describes the properties of space-time, namely its curvature. Therefore it somehow differs
from the other fields which are living on that space-time. If it is really the right approach to
consider space-time curvature as a field living on space-time is debatable. But without this view of
gravity as ``just another field'' there is no reason to quantise gravity at the first place.
As Rosenfeld~\cite{Rosenfeld:1963} points out, the
question if the gravitational field has to be quantised is not for theory to decide but for the experiment.

Contrary to frequent claims, there is no conclusive evidence, neither experimentally nor theoretically, that
gravity cannot be fundamentally classical. In fact, a fully consistent model to describe the interaction of
a classical gravitational and a quantised field was given by Albers \emph{et~al.}~\cite{Albers:2008}. They show that a
two-dimensional version of Nordstr\"om's scalar theory of gravity can be coupled to a quantised massive
scalar field without giving rise to any inconsistencies. This obviously disproves claims that quantisation
of all fields follows as a necessary requirement of mathematical consistency of any quantum theory.
Therefore, this result is also in clear contradiction with a frequently quoted thought experiment by
Eppley and Hannah~\cite{Eppley:1977}.

Eppley and Hannah claim that any theory that couples classical gravity to quantum matter unavoidably
leads to inconsistencies, no matter what nature this coupling is. Their thought experiment is based on
the scattering of a classical gravitational wave with a quantum particle and distinguishes between two situations.

In the first situation, they assume that this scattering acts as a quantum measurement and therefore collapses
the wave-function.
In this scenario, the classical wave---for which the de Broglie relation does not hold---is used to measure the
position of a particle of definite momentum, and in this way it violates the uncertainty principle.
There are at least three problems with this argument. First of all, the analysis by Albers \emph{et~al.}~\cite{Albers:2008} shows that
even for a classical-quantum-coupling the classical field inherits part of the uncertainties of the quantum field
it is coupled to. Thus, it is not true that the classicality of the gravitational wave allows for an arbitrarily precise
position measurement. Second of all, the uncertainty relations are a mere corollary of linear
quantum theory rather than a basic ingredient of the theory and are not experimentally tested in the
situation at hand. They might readily be violated in this case. And finally, Eppley and Hannah
provide a detailed description of their experimental set-up for which it becomes apparent that it is not
realisable---not even in principle---within the parameters of our universe. In particular, the experiment
would require a tremendously massive set of detectors. They would have to be so massive that the detector
arrangement would unavoidably be located within its own
Schwarzschild radius \cite{Mattingly:2006}. Such a coupling of classical gravity to quantum matter, in which
a gravitational measurement collapses the wave-function, therefore yields no obvious contradiction to
fundamental principles of physics.

The semi-classical Einstein equations \eqref{eqn:sce} as well as the \sne~\eqref{eqn:sn}, however,
belong to the second situation which Eppley and Hannah consider: that the scattering of a gravitational
wave leaves the wave-function intact. For this scenario they construct a different type of thought experiment
in which the scattering of a gravitational wave is used to probe the shape of the wave-function (instead of
the expectation values of the observables as usual in quantum mechanics). Making use of this in an EPR-like
set-up opens the possibility of faster-than-light signalling.
One could ask again if the experiment can be conducted, at least in principle. In fact Eppley and Hannah
elaborate much less on this second experiment than on their first one and similar arguments as in the first case
might also render this second thought experiment not feasible.
But at the bottom of this is probably the incompatibility of the realistic interpretation of the wave-function
in the semi-classical equations \eqref{eqn:sce} with the instantaneous, non-local Copenhagen collapse.
This gets even more evident in consideration of the thought experiment allowing superluminous
signalling by means of the \sne\ which we presented in section~\ref{sec:thought-experiment}.
It could very well be, for example, that the problem lies with the collapse of the wave-function, more than
with the way gravity is treated.

There is a second experiment by Page and Geilker~\cite{Page:1981,Page:1982} which is often quoted as an argument for
the necessity of a quantisation of the gravitational field.
In fact, a very similar idea to that of Page and Geilker was already pronounced by Kibble~\cite{Kibble:1981}.
He suggests a black-box in which a quantum
decision-making process is used to create a macroscopic superposition state. According to semi-classical
gravity the gravitational field of such a black-box containing a superposition state would differ from
the gravitational field of one containing a collapsed state.
Page and Geilker, however, claim to actually conduct this experiment and find no evidence for such
a difference of the classical gravitational field. But instead of having a black-box, in their experiment the
so-called ``superposition state''
is created in a purely classical procedure which is completely decoupled
from the decision-making process. Therefore the superposition exists \emph{only} in a no-collapse interpretation
of quantum mechanics and only the inconsistency of such an interpretation with semi-classical gravity
is shown. The authors try to dissolve this obvious flaw in their reasoning by arguing that an instantaneous
collapse would contradict the divergence-free nature of the Einstein equations but this is certainly not a problem
for any form of dynamical description of the collapse.
The conclusion therefore remains what Kibble had already noticed: that this is an indication that
the connection of gravity and quantum mechanics requires understanding the wave-function collapse.

\end{document}